\documentclass[prb,aps,showpacs,floats,twocolumn]{revtex4}
\usepackage{graphicx}

\begin{document}

\title{Mott transitions in three-component Falicov-Kimball model}
\author{Duong-Bo Nguyen and Minh-Tien Tran}
\affiliation{Institute of Physics, Vietnam Academy of Science and Technology, 10 Dao Tan,
Hanoi, Vietnam}

\begin{abstract}
Metal-insulator transitions are studied within a three-component
Falicov-Kimball model which mimics a mixture of one-component and
two-component fermionic particles with  local repulsive
interactions in optical lattices. Within the model the
two-component fermionic particles are able to hop in the lattice,
while the one-component fermionic particles are localized. The
model is studied by using the dynamical mean-field theory with
exact diagonalization. Its homogeneous solutions establish Mott
transitions for both commensurate and incommensurate fillings
between one third and two thirds. At commensurate one third and
two thirds fillings the Mott transition occurs for any density of
hopping particles, while at incommensurate fillings the Mott
transition can occur only for density one half of hopping
particles. At half filling, depending on the repulsive
interactions the reentrant effect of the Mott insulator is
observed. As increasing local interaction of hopping particles,
the first insulator-metal transition is continuous, whereas the
second metal-insulator transition is discontinuous. The second
metal-insulator transition crosses a finite region where both
metallic and insulating phase coexist. At third filling the Mott
transition is established only for strong repulsive interactions.
A phase separation occurs together with the phase transition.
\end{abstract}


\maketitle

\section{Introduction}

Metal-insulator transition (MIT) is a long standing problem in
condensed matter physics and has attracted a lot of attention. The
Mott insulating state is established when conduction electrons
become localized due to the suspension of the double occupation by
electron correlations.\cite{Mott} Most studies on the MIT focus on
electron systems with local Coulomb interaction between the two
spin states of electrons. The achievement of loading ultracold
fermionic atoms in optical lattices has been providing a novel
stage for studying the Mott transitions.\cite{Bloch} Indeed, the
Mott insulating state was realized for $^{40}$K fermionic atoms
with two hyperfine states and repulsive interaction between
them.\cite{Jordens,Schneider} This MIT is similar to the one of
electron systems. Moreover, ultracold fermionic atom mixtures can
be extended to have both large hyperfine multiplet and different
masses. A mixture of single-spin state of $^{40}$K immersed in
two-component fermionic atoms $^6$Li or a mixture of two-component
state $^{171}$Yb and six-component state $^{173}$Yb were
established.\cite{Spiegelhalder,Taie} Such achievements lead to a
possibility of studying the Mott transitions in multi-component
correlation systems. Theoretically, the Mott transitions in
three-component Hubbard models were studied.\cite{Gorelik, Inaba}
Within the dynamical mean-field theory (DMFT) the Mott transition
is observed at commensurate fillings when the local Coulomb
interactions are isotropic.\cite{Gorelik} When the anisotropy of
the local Coulomb interactions is introduced, the Mott transition
is also observed at incommensurate half filling.\cite{Inaba}
However, in these studies, all component particles have the same
masses. The mixtures of ultracold atoms are often established with
mass imbalance. The mass imbalance could affect the MIT. Indeed,
the DMFT of the mass-imbalance mixtures shows that the light
particles may be more affected by correlations than the heavy
ones.\cite{Dao}

In this paper we study the Mott transitions in a three-component
Falicov-Kimball model (FKM). This model can be viewed as a version
of the three-component Hubbard model \cite{Gorelik, Inaba} with
extreme mass imbalance. Within the model, one-component fermionic
particles are extremely heavy and localized. They are immersed in
an optical lattice, where two-component fermionic particles with
light mass can hop. Local repulsive interactions between the
component states of particles are taken into account. The
three-component FKM can be realized by loading mixtures of two
species of fermionic atoms into an optical lattice. One species is
single-spin state heavy atoms and the other is two-component light
atoms. Such ultracold atomic mixtures can be realized by
mixtures of $^{40}$K with $^6$Li, or light $^6$Li or $^{40}$K with
heavy fermionic isotopes of Sr or Yb. When the mixtures are loaded
into an optical lattice, the heavy atoms usually have much lower
tunneling rate than the light atoms. With sufficient lattice
depth, the hopping rate of the heavy atoms can be ignored in
comparison with the one of the light
atoms.\cite{FreericksLemanski} Similar mixtures of two species of
one-component fermionic atoms were also proposed for the
realization of the spinless FKM.\cite{Ates} The three-component
FKM can be considered as a complement of the three-component
Hubbard model, when a strong mass imbalance is realized.
Traditionally, the FKM was introduced to describe the MIT in
rare-earth and transition metal compounds.\cite{FKM} The spinless
version of the FKM has a rich phase diagram, and in particular,
the homogeneous states establish a correlation-driven MIT. At low
temperature charge-density-wave states are
established.\cite{Kennedy} The FKM can be extended to have
additional terms in order to study different physical
phenomena.\cite{Portengen,Tran1,Tran2} Multi-component extensions
of the FKM were also
considered.\cite{Fledderjohann,FreericksZlatic} However, in the
multi-component versions of the FKM, the multiplicity of electron
spins is just a formal generation of the spinless case, and the
Coulomb interactions between the conduction electrons are not
taken into account. In this paper, the three-component FKM  keeps
all local Coulomb interactions between the component states. In
contrast to the spinless case, there is very little knowledge of
the physical properties of the multi-component
FKM.\cite{Jedrzejewski}

The three-component FKM can also be considered as a combination of
the Hubbard model of the two-component hopping particles and the
spinless FKM of the one-component localized particles. At half
filling the Hubbard model exhibits the MIT which is driven by the
Coulomb interaction.\cite{Hubbard} The insulating state is
naturally the Mott insulator,\cite{Mott} where the double
occupation is suspended by the Coulomb interaction. The MIT is
discontinuous.\cite{Georges,Rozenberg} At half filling the
spinless FKM also displays a MIT, although this MIT is
continuous.\cite{FreericksZlatic} When the Coulomb interaction is
large enough, it also prevents the hopping and localized particles
to occupy at the same site. As a result, the insulating state is
established. One may expect that the three-component FKM could
exhibit a rich phase diagram of the MIT. Indeed, in the present
paper we find both continuous and discontinuous MIT. The MIT can
occur not only at half filling, but also at other fillings between
one third and two thirds.

In the present paper we study the three-component FKM by using the
DMFT with exact diagonalization (ED). The DMFT has been widely and
successfully used to study the strongly correlated electron
systems.\cite{Metzner,GKKR} It has been also extensively applied
to study the FKM.\cite{Brandt,FreericksZlatic} The previous
studies of MIT in the three-component Hubbard model were also
based on the DMFT.\cite{Gorelik,Inaba} The advantage of the DMFT
is that it is exact in infinite dimensions and fully captures
local time fluctuations. However, the DMFT loses nonlocal
correlations of finite dimension systems. Within the DMFT we find
that like the three-component Hubbard model,\cite{Inaba} the
three-component FKM exhibits the Mott transition at both
commensurate and incommensurate fillings. However, in contrast to
the three-component Hubbard model, the MIT can occur at any
filling between one third and two thirds. We also find an inverse
MIT, i.e. the transition from insulator to metal when particle
correlations increase. The inverse MIT can occur only in the
region of large repulsive interactions between localized and
hopping particles. At half filling, a reentrant effect of the Mott
insulator is observed. With increasing Coulomb interactions, the
system first stays in the insulator region, then it becomes
metallic, and finally goes back to the insulating phase. The first
transition is continuous, while the second one is discontinuous.
At the second MIT, we observe a finite region where both metallic
and insulating phases coexist, like the MIT in the single-band
Hubbard model.\cite{Georges,GKKR} However, in contrast to the
single-band Hubbard model, the MIT in the three-component FKM is
due to a superposition of two Kondo resonances that are shifted
from the Fermi level. At one third and two thirds fillings, the
Mott transition is established only for strong repulsive
interactions and a phase separation occurs together with the phase
transition.

The present paper is organized as follows. In Sec. II we present
the three-component FKM and its DMFT. The general filling
conditions for the MIT are presented Sec. III. In this section we
also study the MIT at half and third filling in details. Finally,
the conclusion is presented in Sec. IV.

\section{Three-component Falicov-Kimball model and its dynamical mean field theory}

We consider a three-component FKM that describes a  mixture of
one-component heavy fermionic particles and two-component light
fermionic particles loaded in an optical lattice. The heavy
particles are localized, whereas the light particles can hop in the
lattice. Its Hamiltonian reads
\begin{eqnarray}
H &=& -t\sum_{<i,j>,\sigma} c^{\dagger}_{i\sigma} c_{j\sigma} + U_{cc}\sum_{i}
c^{\dagger}_{i\uparrow} c_{i\uparrow} c^{\dagger}_{i\downarrow}
c_{i\downarrow} \nonumber \\
&& + E_{f} \sum_{i} f^{\dagger}_{i} f_{i} +
U_{cf} \sum_{i\sigma} f^{\dagger}_{i} f_{i}  c^{\dagger}_{i\sigma} c_{i\sigma}
,  \label{ham}
\end{eqnarray}
where $c^{\dagger}_{i\sigma}$ ($c_{i\sigma}$) is the creation
(annihilation) operator for fermionic particle with hyperfine
multiplet (or spin) $\sigma$ at site $i$. $\sigma$ takes two
values $\pm 1$. $t$ is the hopping parameter of the two-component
fermionic particles, and we  take into account only the hopping
between nearest neighbor sites. $U_{cc}$ is the local Coulomb
interaction between the two component states of those particles.
$f^{\dagger}_{i}$ ($f_{i}$) is the creation (annihilation)
operator for one-component (or spinless) fermionic particle at
site $i$. $U_{cf}$ is the local Coulomb interaction between the
two-component particles and one-component particles. The
one-component particles do not move, and its energy level is
$E_{f}$. $E_f$ can also be considered as the chemical potential of
the localized particles and controls the filling of the
localized particles $n_f=\sum_{i} \langle f^{\dagger}_{i} f_i
\rangle /N$, where $N$ is the number of lattice sites. A common
chemical potential $\mu$ is introduced to control the total
particle filling $n=\sum_{\sigma} n_{c\sigma} + n_f$, where
$n_{c\sigma}=\sum_{i} \langle c^{\dagger}_{i\sigma} c_{i\sigma}
\rangle /N$. The three-component FKM has two well-known limiting
cases. When $U_{cf}=0$ the two-component and one-component
 particles are completely decoupled. The Hamiltonian of
  the two-component particles is just
  the single-band Hubbard model.\cite{Hubbard}
  Within
   the DMFT, the Mott insulating phase is established
   at half filling and sufficient large local Coulomb interactions.
   The MIT is of first order.\cite{Georges,GKKR} When $U_{cc}=0$ the model
   in Eq. (\ref{ham})
   is equivalent to the spinless
FKM.\cite{FKM} Within the DMFT its homogeneous phase also displays
a MIT.\cite{FreericksZlatic} However, the metallic phase breaks
the Fermi liquid  theory due to the pinning of the chemical
potential at the localized particle level.\cite{Si} The MIT is
continuous. The model in Eq. (\ref{ham}) can also viewed as a
simplified version of the three-component Hubbard model
\cite{Gorelik,Inaba} with strong mass imbalance, where the
particles of one specified component are heavy and localized. The
three-component FKM can be realized by  loading mixtures of
one-component heavy fermionic particles and two-component light
fermionic particles into an optical lattice. In a deep enough
lattice, the hopping rate of the heavy particles is much lower
than the one of the light particles, and it can be ignored.
In the model in Eq. (\ref{ham}) we also neglect the trapping potentials. They must be included when realistic MIT is observed and compared with the theoretical calculations.

We study the three-component FKM by using the DMFT. Within the DMFT the self energy
is a local function of frequency. The Green function of the two-component particles
reads
\begin{equation}
G(\mathbf{k},i\omega_n) = \displaystyle\frac{1}{i\omega_n+\mu-\varepsilon_{\mathbf{k}}-\Sigma(i\omega_n)},
\end{equation}
where $\omega_n$ is the Matsubara frequency,
$\varepsilon_{\mathbf{k}}$ is the dispersion of the two-component
particles, and $\Sigma(i\omega_n)$ is the self  energy. Here we
are interested in the homogeneous phase for all particle
components, hence the $\sigma$ index as well as the site index of
the self energy are omitted. The self energy is determined from the dynamics of a
single two-component particle embedded in an effective self
consistent medium. The action of the effective system reads
\begin{eqnarray}
{\cal S}_{{\rm imp}} =  - \int\limits_{0}^{\beta}
\int\limits_{0}^{\beta} d\tau d\tau' \sum_{\sigma}
c^{\dagger}_{\sigma}(\tau) {\cal{G}}^{-1}(\tau-\tau')
c_{\sigma}(\tau') \nonumber \\
+ U_{cc} \int\limits_{0}^{\beta} d\tau (c^{\dagger}_{\uparrow}
c_{{\uparrow}} c^{\dagger}_{\downarrow} c_{\downarrow})(\tau)
 + (E_{f}-\mu) \int\limits_{0}^{\beta} d\tau f^{\dagger}(\tau)
f(\tau) \nonumber \\
+ U_{cf} \int\limits_{0}^{\beta} d\tau
\sum_{\sigma} (c^{\dagger}_{\sigma} c_{\sigma} f^{\dagger}
f)(\tau), \label{action}
\end{eqnarray}
where ${\cal{G}}(\tau)$ is a Green function which represents the effective
medium. It relates to the self energy and the local Green function by the
Dyson equation
\begin{equation}
{\cal{G}}^{-1}(i\omega_n) = G^{-1}(i\omega_n) + \Sigma(i\omega_n) .
\label{dyson}
\end{equation}
Here, the local Green function is
\begin{equation}
G(i\omega_n) = \int d\varepsilon \rho_{0}(\varepsilon)
\displaystyle\frac{1}{i\omega_n+\mu-\Sigma(i\omega_n)-\varepsilon},
\label{local}
\end{equation}
where $\rho_{0}(\varepsilon)=\sum_{\mathbf{k}}
\delta(\varepsilon-\varepsilon_{\mathbf{k}})$ is the bare density
of states (DOS). Without loss of generality, we use the
semicircular DOS
\begin{equation}
\rho_0(\varepsilon) = \frac{2}{\pi D^2} \sqrt{D^2-\varepsilon^2} ,
\end{equation}
where $D$ is the half bandwidth. We will use $D$ as the unit of energy.
With the semicircular DOS, from Eqs. (\ref{dyson})-(\ref{local})  we obtain \cite{GKKR}
\begin{equation}
{\cal G}^{-1}(i\omega_n) = i\omega_n + \mu - \frac{D^2}{4} G(i\omega_n) .
\end{equation}
 One can notice that the occupation number of localized particles in the effective action
 in Eq. (\ref{action}) is conserved. It  can take only   two values: $0$ and $1$.
  Therefore, the partition function of the effective action can be evaluated
independently in the sectors of $f^{\dagger}f=0,1$. We obtain
\begin{eqnarray}
{\cal Z}_{{\rm imp}} &=& {\rm Tr}_{f} \int  \prod_{\sigma}
{\cal D} c^{\dagger}_{\sigma} {\cal D} c_{\sigma} e^{-{\cal S}_{{\rm imp}}} \nonumber \\
&=& \int \prod_{\sigma} {\cal D} c^{\dagger}_{\sigma} {\cal D} c_{\sigma}
e^{-{\cal S}_{0}[c^{\dagger}_{\sigma},c_{\sigma} ]} \nonumber \\
&& +e^{-\beta(E_f-\mu)} \int \prod_{\sigma} {\cal D} c^{\dagger}_{\sigma} {\cal D} c_{\sigma}
e^{-{\cal S}_{1}[c^{\dagger}_{\sigma},c_{\sigma} ]} \nonumber \\
&=& {\cal Z}_{0} + e^{-\beta(E_f-\mu)} {\cal Z}_{1}
,
\end{eqnarray}
where
\begin{eqnarray}
{\cal Z}_m = \int \prod_{\sigma} {\cal D} c^{\dagger}_{\sigma}
{\cal D} c_{\sigma} e^{-{\cal S}_m[c^{\dagger}_{\sigma},c_{\sigma} ]} \mbox{\hspace{1.5cm}}, \label{zm} \\
 {\cal S}_{m}[c^{\dagger}_{\sigma},c_{\sigma} ] =
-\int\limits_{0}^{\beta} \int\limits_{0}^{\beta} d\tau d\tau'
\sum_{\sigma} c^{\dagger}_{\sigma}(\tau) [{\cal{G}}^{-1}(\tau-\tau') \nonumber \\
-m U_{cf}\delta(\tau-\tau') ] c_{\sigma}(\tau')
+ U_{cc} \int\limits_{0}^{\beta} d\tau
(c^{\dagger}_{\uparrow} c_{{\uparrow}} c^{\dagger}_{\downarrow}
c_{\downarrow})(\tau) ,
\label{actionm}
\end{eqnarray}
with $m=0,1$. In contrast to the spinless  case of the
FKM,\cite{FreericksZlatic,Brandt} the partition function in Eq.
(\ref{zm}) cannot be evaluated analytically. It has the same form
of the partition function of a single site of the Hubbard model
embedded in an effective mean field medium.\cite{GKKR} Suppose we
can solve the effective action in Eq. (\ref{actionm}) and obtain
the Green function
\begin{eqnarray}
\lefteqn{
G_{m}(i\omega_n) = - \displaystyle\frac{\delta \ln {\cal Z}_m}
{\delta \lambda(i\omega_n)} } \nonumber \\
&&= \displaystyle\frac{1}{i\omega_n + \mu -\lambda(i\omega_n)- m U_{cf} -\Sigma_m(i\omega_n) },
\end{eqnarray}
where $\lambda(i\omega_n)=i\omega_n+\mu-{\cal{G}}^{-1}(i\omega_n)$, and
$\Sigma_{m}(i\omega_n)$ is the corresponding self energy due to the local Coulomb
interaction $U_{cc}$. Once the self energy $\Sigma_{m}(i\omega_n)$ is known, the
Green function of the effective action in Eq. (\ref{action}) can also be  determined. We obtain
\begin{eqnarray}
\lefteqn{
G_{{\rm imp}}(i\omega_n) = - \displaystyle\frac{\delta \ln {\cal Z}_{{\rm imp}}}
{\delta \lambda(i\omega_n)} }  \nonumber \\
&& = \displaystyle\frac{w_0}{i\omega_n + \mu -\lambda(i\omega_n)-\Sigma_0(i\omega_n) } \nonumber \\
&& +\displaystyle\frac{w_1}{i\omega_n + \mu  -\lambda(i\omega_n)-U_{cf}-\Sigma_1(i\omega_n) },
\label{gimp}
\end{eqnarray}
where
\begin{eqnarray}
w_m = \frac{e^{-m \beta (E_f-\mu)} {\cal Z}_m}{{\cal Z}_{{\rm imp}}} .
\end{eqnarray}
One can notice that $w_0+w_1=1$, and $w_1$ is just the filling of
localized particles. Basically, the Green function of
the effective action in Eq. (\ref{gimp}) has the same structure of
the one of the spinless case,\cite{FreericksZlatic,Brandt} except of
the additional self energy $\Sigma_m(i\omega_n)$ due to the local
Coulomb interaction between the hopping particles. The self
consistent condition of the DMFT requires that the Green function
obtained from the effective action in Eq. (\ref{action}) must
coincide with the local Green function in Eq. (\ref{local}), i.e,
\begin{equation}
G_{{\rm imp}}(i\omega_n) = G(i\omega_n) .
\end{equation}
Now we obtain the complete self consistent system of equations for
the Green function. It can be solved numerically by
iterations.\cite{GKKR} The most  time consuming part is the
solving of the action ${\cal S}_m$ in Eq. (\ref{actionm}). We
apply ED to solve it.\cite{GKKR,Krauth} The action in Eq.
(\ref{actionm}) describes the dynamics of an impurity with the
repulsive interaction coupling with a conduction bath. It is
essentially equivalent to the Anderson impurity model
\begin{eqnarray}
H_{m} &=& (\mu-m U_{cf}) \sum_{\sigma} c^{\dagger}_{\sigma} c_{\sigma} +
U_{cc} c^{\dagger}_{\uparrow} c_{\uparrow} c^{\dagger}_{\downarrow}
c_{\downarrow} \nonumber \\
&&+ \sum_{p,\sigma} V_{p} a^{\dagger}_{p\sigma} c_{\sigma} + {\rm h.c.}
+ \sum_{p,\sigma} E_{p} a^{\dagger}_{p\sigma} a_{p\sigma} ,
\label{anderson}
\end{eqnarray}
where $a^{\dagger}_{p\sigma}$ ($a_{p\sigma}$) is the creation
(annihilation) operator which represents the conduction bath with
the energy level $E_p$. $V_p$ is the coupling of the conduction
bath with the impurity. The connection between  the Anderson
impurity model in Eq. (\ref{anderson}) and the action in Eq.
(\ref{actionm}) is the following identity relation of the bath
parameters \cite{GKKR,Krauth}
\begin{equation}
\sum_{p} \frac{V_p^2}{i\omega_n - E_p} = \lambda(i\omega_n) .
\end{equation}
To carry out the ED, the conduction bath is limited to finite $n_s$ orbits ($p=1,2,...,n_s$).
Then the bath Green function ${\cal G}^{-1}(i\omega_n)$ is approximated by
\begin{equation}
{\cal G}_{n_s}^{-1}(i\omega_n) = i\omega_n+\mu-m U_{cf} - \lambda_{n_s}(i\omega_n),
\end{equation}
where $\lambda_{n_s}(i\omega_n)=\sum_{p=1}^{n_s}V_p^2/(i\omega_n - E_p)$.
The bath parameters are determined from minimization of the distance $d$
between ${\cal G}^{-1}(i\omega_n)$ and ${\cal G}_{n_s}^{-1}(i\omega_n)$
\begin{equation}
d = \frac{1}{M+1} \sum_{n=0}^{M}
\omega_n^{-k} |{\cal G}^{-1}(i\omega_n)-{\cal G}_{n_s}^{-1}(i\omega_n) |^2 ,
\end{equation}
where $M$ is a large upper cutoff of the Matsubara
frequencies.\cite{GKKR,Krauth} The parameter $k$ is introduced to
enhance the importance of low Matsubara frequencies in the
minimization procedure. In particular, we take $k=1$ in the
numerical calculations. We use the LISA package to find the bath
parameters as well as to perform the ED.\cite{GKKR} When the bath
parameters are determined, we calculate the Green function of the
Anderson impurity model in Eq. (\ref{anderson}) by
ED.\cite{GKKR,Krauth} In particular, the numerical results
presented in the following sections are obtained by ED with the
bath size $n_s=4$. For the single-band Hubbard model, the bath
size $n_s=4$ already gives good quantitative results at low
temperature in comparison with the Monte Carlo
simulations.\cite{GKKR} We have also checked some numerical
results by comparing them with the ones calculated with the bath
size $n_s=5$. A careful analysis of the ED DMFT shows that two
bath levels per impurity orbit usually give adequate
results.\cite{Liebsch} When the ED is performed, we compute the
Green function and its self energy at temperature $T=0.01$. We
consider the grand canonical ensemble and take the filling of
localized particles $n_f$ as an input parameter, instead of the
energy level $E_f$. As we have noticed $w_1=n_f$ that simplifies
the numerical calculations.

\section{Metal-insulator transitions}
\subsection{General filling case}

\begin{figure}[t]
\includegraphics[width=0.47\textwidth]{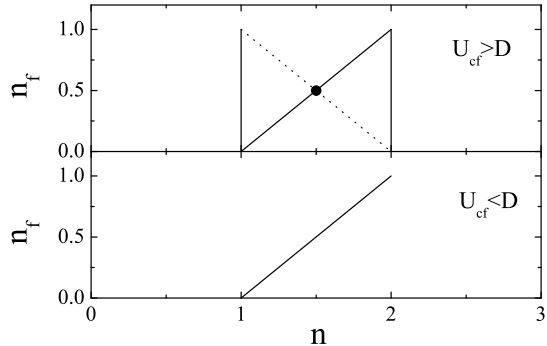}
\caption{Diagram of particle fillings for the MIT. The solid lines
show the
 filling values of $(n,n_f)$ at which the MIT can occur,
and the dotted line shows the filling values of $(n,n_f)$ for the
occurrence of the inverse MIT. The black point indicates the
filling value $(n=3/2,n_f=1/2)$ for the reentrant effect of the
MIT.} \label{fig5}
\end{figure}

As we have mentioned in the previous section, the three-component
FKM has two well-known limiting cases. When $U_{cf}=0$, the
Hamiltonian of the hopping particles is just the single-band
Hubbard model. At  half filling an MIT occurs at low temperature
and this transition is of first order. There is a finite range of
the Coulomb interaction $U_{c1}<U_{cc}<U_{c2}$ where both metallic
and insulating phases coexist.\cite{Georges,GKKR} When $U_{cc}=0$
the three-component FKM is equivalent to the spinless one. For the
homogeneous phase a MIT also occurs at half filling. In contrast
to the single-band Hubbard model, this transition is continuous at
the transition point $U_{cf}^{c}=D$.\cite{FreericksZlatic}  The
situation may completely change when both $U_{cc}$ and $U_{cf}$
terms take effect. First, we summarize the particle filling
conditions at which a MIT can occur. In Fig. \ref{fig5} we plot
the particle filling diagram for MIT. The lines in this figure
show the filling values of $n$ and $n_f$ at which the MIT can
occur. Outside these lines no MIT is observed. There is also a
possibility of the transition from insulator to metal  when the
driven interaction increases. We refer the transition as an
inverse MIT. In Fig. \ref{fig5} the particle filling condition for
the inverse MIT is presented by the dotted line. The particle
filling diagram is different when $U_{cf}>D$ and $U_{cf}<D$ as
shown in Fig. \ref{fig5}. In the homogeneous phase $n=2
n_{c\sigma}+n_f$. The MIT can occur when $1<n<2$. In this case
always $n_{c\sigma}=1/2$ independently whether $U_{cf}>D$ or
$U_{cf}<D$. Therefore, the MIT is driven only by $U_{cc}$ like in
the single-band Hubbard model. The MIT can occur only at
$n_{c\sigma}=1/2$, where every lattice site is occupied by one
light particle, and their strong local Coulomb interaction does not
allow the double occupation of the light particles. The MIT may be
interpreted as a species selective MIT, where the two-component
light particles are localized due to their Coulomb interaction.
The number of heavy particles, as well as their Coulomb
interaction with the light particles are irrelevant to this MIT.
One may also expect that this MIT is of first order. In Fig.
\ref{fig5} one can also see that the MIT can also occur at $n=1$
or $n=2$ for $U_{cf}>D$. In this case the fillings of
two-component and one-component particles can be arbitrary, but
their total filling is always one third or two thirds. The total
filling is commensurate with the number of particle components. At
$n=1$, every lattice site is occupied by one particle of any
component. The Coulomb interactions (both $U_{cc}$ and $U_{cf}$)
suppress any double occupation of particles. At $n=2$, the same
scenario happens, but with holes instead of particles in the case
$n=1$. We refer this MIT as a commensurate MIT. The same MIT occurs
in the three-component Hubbard model.\cite{Gorelik,Inaba} In the
commensurate MIT, the light and heavy particles play equal roles.
The Coulomb interactions prevent any double occupation of
particles. In addition to the species selective and commensurate
MIT, we also observe the inverse MIT. The inverse MIT can occur
when $n_{c\sigma}+n_{f}=1$ and $U_{cf}>D$. These conditions are
similar to the ones in the spinless FKM.\cite{FreericksZlatic} In
this case, every lattice site is occupied by one light or one
heavy particle. The Coulomb interaction between the light and
heavy particles does not allow them to occupy the same lattice
site. Since when $U_{cc}=0$ the system is in the insulating phase,
one may expect that with increasing $U_{cc}$ the system will
transform to the metallic phase. When both the MIT and the inverse
MIT occur, a reentrant effect of MIT could occur. It happens at
half filling $n=3/2$ and $n_f=1/2$.

\begin{figure}[t]
\includegraphics[width=0.47\textwidth]{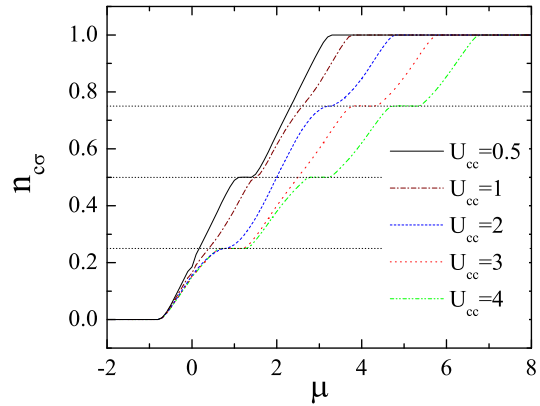}
\caption{(Color online) The hopping particle filling $n_{c\sigma}$
as a function of the chemical potential $\mu$ at $n_{f}=1/2$ for
different values of $U_{cc}$ and fixed $U_{cf}=2$. The horizontal
dotted lines show $n_{c\sigma}=1/4$, $n_{c\sigma}=1/2$, and
$n_{c\sigma}=3/4$ ($T=0.01$, $D=1$).} \label{fig1}
\end{figure}

Next, we present the numerical results to verify the diagram
plotted in Fig. \ref{fig5}. For this purpose we study the
dependence of the filling of light particles $n_{c\sigma}$ on the
chemical potential $\mu$ at a given value of $n_f$. In the
insulating phase when the chemical potential lies in the band gap
the filling $n_{c\sigma}$ does not change. The graphics of the
filling $n_{c\sigma}$ as a function of the chemical potential must
show a plateau when the chemical potential lies in the band gap.
In Fig. \ref{fig1} we present the numerical results of
$n_{c\sigma}$, which are obtained by solving the self-consistent
equations of the DMFT plus ED, as a function of the chemical
potential at $n_{f}=1/2$ and $U_{cf}=2$. Since $U_{cf} > D$, it is
clear that for small values of $U_{cc}$ the system is insulating
at $n_{c\sigma}=1/2$. In this insulating phase each lattice site
is occupied by one light or one heavy particle, and the Coulomb
interaction between them prevent their double occupation. In Fig.
\ref{fig1} it is shown by the plateau of the line with
$U_{cc}=0.5$. With increasing $U_{cc}$ the system changes to a
metallic phase, where the graphics of $n_{c\sigma}$ does not
exhibit any plateau. It is the inverse MIT, which occurs when
$n_{c\sigma}+n_{f}=1$. With further increasing $U_{cc}$, the
graphics of $n_{c\sigma}$ begins to develop two plateaus at
$n_{c\sigma}=1/4$ and $n_{c\sigma}=3/4$, which correspond to the
total filling $n=1$ and $n=2$. These plateaus signal the
insulating phase at one third and two thirds of the total filling.
The corresponding MIT is commensurate. For sufficient large
$U_{cc}$, an additional plateau appears at $n_{c\sigma}=1/2$ that
indicates the insulating phase at $n_{c\sigma}=1/2$. Therefore at
$n_{c\sigma}=1/2$ there is a reentrance of the insulating phase.
However, in contrast to the insulating phase at small values
$U_{cc}$, in this insulating phase, each lattice site is occupied
by one light particle and the Coulomb interaction of the light
particles prevents the double occupation of the light particles,
because the same MIT also occurs at small $U_{cf}$. This species
selective MIT does not depend on $U_{cf}$ as well as the number of
heavy particles.

\begin{figure}[t]
\includegraphics[width=0.45\textwidth]{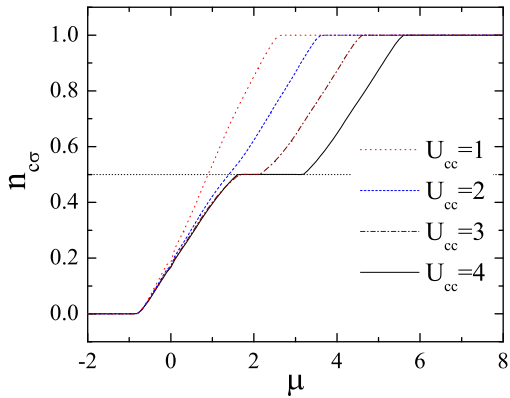}
\caption{(Color online) The hopping particle filling $n_{c\sigma}$
as a function of the chemical potential $\mu$ at $n_{f}=1/2$ for
different values of $U_{cc}$ and fixed $U_{cf}=0.8$. The
horizontal dotted line shows $n_{c\sigma}=1/2$ ($T=0.01$, $D=1$).}
\label{fig2}
\end{figure}

In Fig. \ref{fig2} we plot the filling $n_{c\sigma}$ at
$U_{cf}=0.8$ and $n_{f}=1/2$. In this case $U_{cf}<D$, thus the
system is metallic when $U_{cc}=0$.  Weak correlations of light
particles cannot drive the system out the metallic state. However,
with further increasing $U_{cc}$, the particle correlations drive
the system into insulating state. It is shown in Fig. \ref{fig2}
by the appearance of plateau in the graphics of $n_{c\sigma}$. The
MIT occurs at $n_{c\sigma}=1/2$, and it is the species selective
MIT. For $U_{cf}<D$ we do not observe any MIT at one third or two
thirds of the total filling. The correlations between the light
and heavy particles is not strong enough to prevent them to occupy
the same site. Consequently, the commensurate MIT cannot be
established.

We have checked these about analyzing results for both the cases
$U_{cf}>D$ and $U_{cf}<D$ at general filling $n_f=p/q$, where $p,
q$ are integer ($p<q$). In summary, the three-component FKM
exhibits different MIT depending on the particle fillings and the
Coulomb interactions. The MIT can be species selective,
commensurate or inverse transitions. The species selective MIT
occurs at $n_{c\sigma}=1/2$, while the commensurate MIT occurs
only at one third or two thirds of total filling. The inverse MIT
can occur at $n_{c\sigma}+n_{f}=1$ and $U_{cf}>D$. These MIT could
be observed by loading mixtures of light and heavy fermionic atoms
into an optical lattice with sufficient lattice depth. By
measuring the double occupations among the light particles,
between the light and heavy particles, one could detect these MIT.

The half and one third fillings are two special cases of
the MIT. In the next two subsections we will study the cases in
detail. We will analyze the MIT through the self energy and the
DOS of the light particles. The experimental observations of
spectra of the light particles in optical lattices are
challenging.

\subsection{Half filling case}

\begin{figure}[t]
\includegraphics[width=0.47\textwidth]{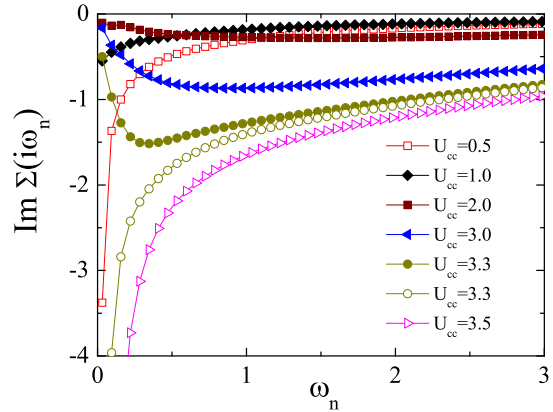}
\caption{(Color online) The imaginary part of the light particle
self energy at half filling for different values of $U_{cc}$ and
fixed $U_{cf}=1.6$. ($T=0.01$, $D=1$). } \label{fig6}
\end{figure}


\begin{figure}[t]
\includegraphics[width=0.47\textwidth]{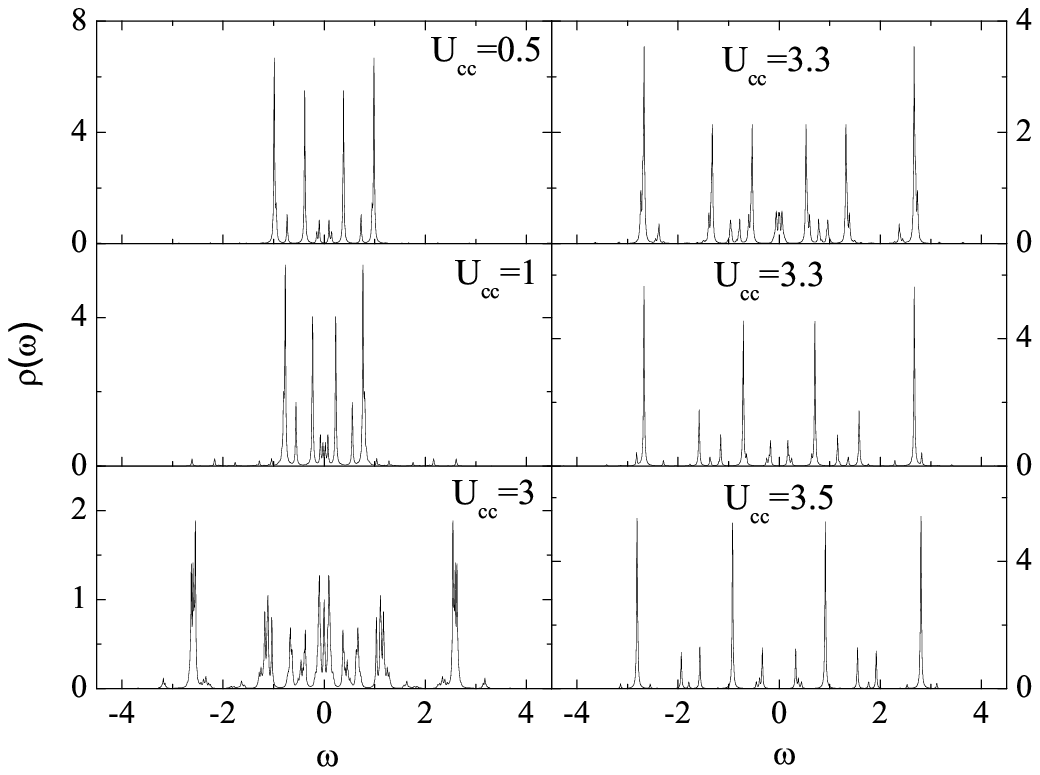}
\includegraphics[width=0.47\textwidth]{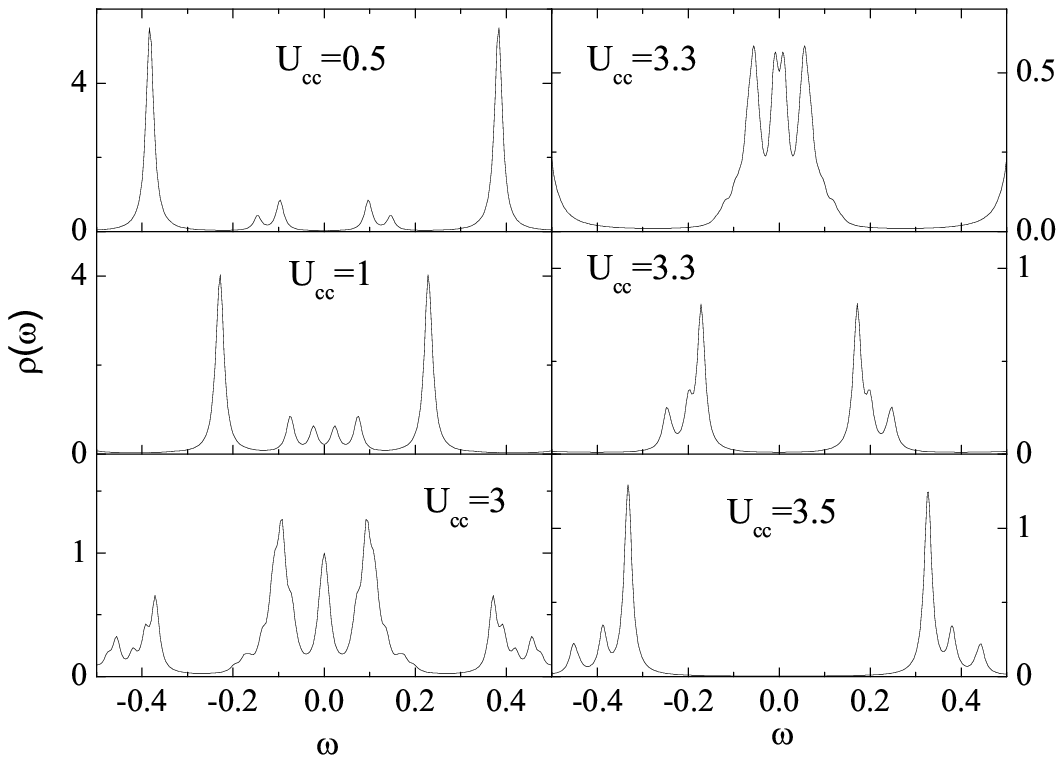}
\caption{The DOS of the light particles at half filling for
different values of $U_{cc}$ and $U_{cf}=1.6$. Top panel: the DOS
at the full energy scale. Bottom panel: the DOS is focused at low
energies. ($T=0.01$, $D=1$).} \label{fig7}
\end{figure}


In this subsection we study the MIT at half filling in details.
One can see in Fig. \ref{fig5} that at half filling ($n=3/2$) the
MIT can occur only for $n_{c\sigma}=n_{f}=1/2$. It turns out that
$\mu=(U_{cc}+U_{cf})/2$ with the particle-hole symmetry. In Fig.
\ref{fig6} we present the numerical results of the self energy
${\rm Im} \Sigma(i\omega_n)$ of the light particles for different
values of $U_{cc}$ and a fixed $U_{cf}$. The behavior of ${\rm Im}
\Sigma(i\omega_n)$ at low frequencies indicates the conduction
properties of the system. When ${\rm Im} \Sigma(i\omega_n)$
diverges as $\omega_n \rightarrow 0$, the system is insulating,
while when ${\rm Im} \Sigma(i\omega_n) \rightarrow 0$ as $\omega_n
\rightarrow 0$, the system is metallic. If $\lim_{\omega_n
\rightarrow 0}{\rm Im} \Sigma(i\omega_n)$  is finite, the system
is still metallic, but it does not obey the Fermi liquid
properties. Since in Fig. \ref{fig6}, $U_{cf} >D$, one may expect
that the system is insulating for small values of $U_{cc}$.
Indeed, in Fig. \ref{fig6} we can see that ${\rm Im} \Sigma(i\omega_n)$
diverges as $\omega_n \rightarrow 0$ for $U_{cc}=0.5$ which
indicates the insulating state of the system. With increasing
$U_{cc}$, ${\rm Im} \Sigma(i\omega_n)$ stops to diverge as
$\omega_n \rightarrow 0$. It shows that the system becomes
metallic. However, ${\rm Im} \Sigma(i\omega_n)$ may tend to a
finite value when $\omega_n \rightarrow 0$. In Fig. \ref{fig6}
this behavior is shown by the self energy with $U_{cc}=1$ and
$U_{cc}=2$. In this regime, the finite value of ${\rm Im}
\Sigma(i\omega_n)$ when $\omega_n \rightarrow 0$ becomes smaller
with increasing $U_{cc}$. With further increasing $U_{cc}$, ${\rm
Im} \Sigma(i\omega_n)$ has a tendency of decreasing its value  at
$\omega_n \rightarrow 0$ to zero. This behavior reminisces the
Fermi liquid properties. It is shown in Fig. \ref{fig6} by the
self energy with $U_{cc}=3$. With further increasing $U_{cc}$,
${\rm Im} \Sigma(i\omega_n)$ diverges as $\omega_n \rightarrow 0$,
and the system falls into the insulating phase regime again.
However, between the metallic and insulating phase we detect a
finite region, where both the metallic and insulating solutions
coexist. This feature is similar to the MIT in the single-band
Hubbard model.\cite{Georges,GKKR} In Fig. \ref{fig6} we present
both metallic and insulating self energies in this region
($U_{cc}=3.3$). In the coexistent region distinct metallic and
insulating phases do not exist. It is similar to the critical
point of classical gases, above which no phase boundaries between
vapor and liquid phases exist. From the behaviors of the self
energy, we observe the reentrance of the insulating phase in the
region of $U_{cf}>D$. With increasing $U_{cc}$ from zero value,
the system first stays in the insulating phase, then the Coulomb
interactions drive it into the metallic phase, and finally the
system goes back to the insulating phase. The later MIT is first
order, and it crosses a finite region where both the metallic and
insulating phases coexist.

The above analyzed phase transitions can also be seen from the
behavior of the DOS of the light particles. In Fig. \ref{fig7} we
plot the DOS $\rho(\omega)=- {\rm Im} G(\omega+i\eta)/\pi$ for
different values of $U_{cc}$ and fixed $U_{cf}$. In ED we can
compute the Green function for both Matsubara and real
frequencies. In the numerical calculations we take $\eta=0.01$ for
broadening the width of the delta functions. When the DOS shows a
gap at the position of the chemical potential ($\omega=0$), the
system is insulating. If the DOS is finite at $\omega=0$ the
system is metallic. Fig. \ref{fig7} shows a MIT between
$0.5<U_{cc}<1$ at $U_{cf}=1.6$. For small values of $U_{cc}$,  the
DOS exhibits a gap at the Fermi energy. With increasing $U_{cc}$
the gap becomes smaller, and it disappears at the transition point
$U_c$. When $U_{cc}>U_{c}$, the DOS exhibits a pseudogap at
$\omega=0$. The pseudogap is developed from the gap of the
insulating phase. In Fig. \ref{fig7}, this behavior is shown by
the DOS with $U_{cc}=1$. The phase with pseudogap corresponds to
the case of finite value of the imaginary part of the self energy
at zero energy. With further increasing $U_{cc}$, the DOS exhibits
a group of narrow peaks around $\omega=0$. In Fig. \ref{fig7} it
is shown by the DOS with $U_{cc}=3$. In the single-band Hubbard
model, the group of narrow peaks at $\omega=0$ represents the
Kondo resonance.\cite{GKKR,Krauth} However, due to the finite size
effect and the discreteness of the ED, the Kondo resonance does
not appear at the full scale. Nevertheless, the appearance of the
group of narrow peaks at $\omega=0$ can be interpreted as a signal
of the Kondo resonance. However, in contrast to the single-band
Hubbard model, the local Green function here is a superposition of
two local subband Green functions $G_{m}(\omega)$ as one can see
from Eq. (\ref{gimp}). Each local subband Green function is
described by an Anderson impurity with the Coulomb interaction
$U_{cc}$ embedded in a conduction bath. The energy level of the
Anderson impurity is $\mu -m U_{cf} = (U_{cc} \pm U_{cf})/2$. The
Anderson impurity is symmetric when its energy level is placed at
$U_{cc}/2$.\cite{Hewson,Horvatic} Since $U_{cf}> 0$, the Anderson
impurity is asymmetric. Therefore, the Kondo resonance that
appears in the local subband DOS, must be shifted from the Fermi
level.\cite{Hewson,Horvatic} Only for the symmetric Anderson
impurity the Kondo resonance peaks at the Fermi level. 
Consequently, the group of narrow peaks at $\omega=0$ represents
the superposition of the two shifted Kondo peaks. It can be
considered as an asymmetric Kondo splitting. Experimental
observations of the asymmetric Kondo splitting in the spectra of
the light particles are challenging. With tunable value of
$U_{cf}$, experiments could observe the asymmetric Kondo splitting
in the spectra of light particles and the splitting width
increases with increasing $U_{cf}$. At $U_{c2}$ the group of
narrow peaks disappears, and the system becomes insulating.
However, as in the single band Hubbard model, the insulating
solution exists until $U_{c1}<U_{c2}$. In the region
$U_{c1}<U_{cc}<U_{c2}$ both metallic and insulating solutions
coexist. Thus, the second MIT is of first order. This MIT solely
deals with the light particles, because it exists even when
$U_{cf}$ is small. The transition is the species selective MIT. In
the insulating phase each lattice site is occupied by one light
particle, and the Coulomb interaction prevents the double
occupation of the light particles. Experiments could detect the
existence of the coexistent region by measuring the double
occupation of the light particles. Across the region, the double
occupation must exhibit a jump in its value. The MIT mechanisms of
the two MIT are different. In the first inverse MIT, the pseudogap
of the metallic phase is developed from the gap of the insulating
phase, whereas in the second species selective MIT, the phase
transition occurs due to the collapse of the asymmetric Kondo
splitting.

\begin{figure}[t]
\includegraphics[width=0.47\textwidth]{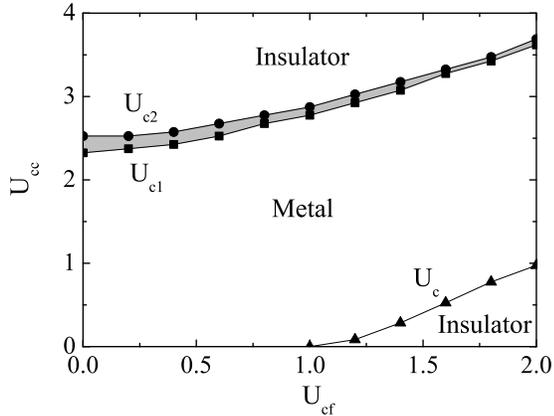}
\caption{Phase diagram at half filling $n_{c\sigma}=n_f=1/2$.  In
the grey area, both metallic and insulating phase coexist. }
\label{fig10}
\end{figure}

We have also analyzed the MIT through the self energy and the DOS
for the case $U_{cf}<D$. In this case the MIT is species
selective, and has the same features as the one in the case
$U_{cf}>D$, which we have discussed above. We summarize the phase
transitions in the phase diagram plotted in Fig.
\ref{fig10}. In the region $U_{cf}>D$, there are two MITs. One is
inverse MIT and the other is species selective MIT. At the first
transition ($U_c$), particle correlations drive the system from
insulating to metallic phase, while at the second one, particle
correlations drive the system from metallic to insulating phase. A
finite region of coexistence of the metallic and insulating phase
exists between the species selective MIT.  The critical value
$U_c$ of the inverse MIT increases with $U_{cf}$. It vanishes at
$U_{cf}=D$. Hence, in the region $U_{cf}<D$ the reentrant effect
of the insulating phase is absent. The phase diagram is consistent
with the analysis of the phase transitions observed from
dependence of $n_{c\sigma}$ on the chemical potential which have
been presented in the previous subsection.


\subsection{Third filling case}

\begin{figure}[t]
\includegraphics[width=0.4\textwidth]{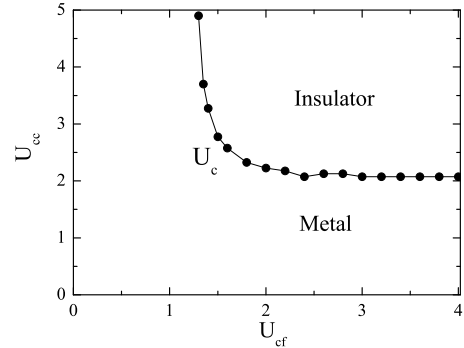}
\caption{Phase diagram for third filling $n_{c\sigma}=n_f=1/3$.
($T=0.01$, $D=1$).} \label{fig14}
\end{figure}

\begin{figure}[t]
\includegraphics[width=0.4\textwidth]{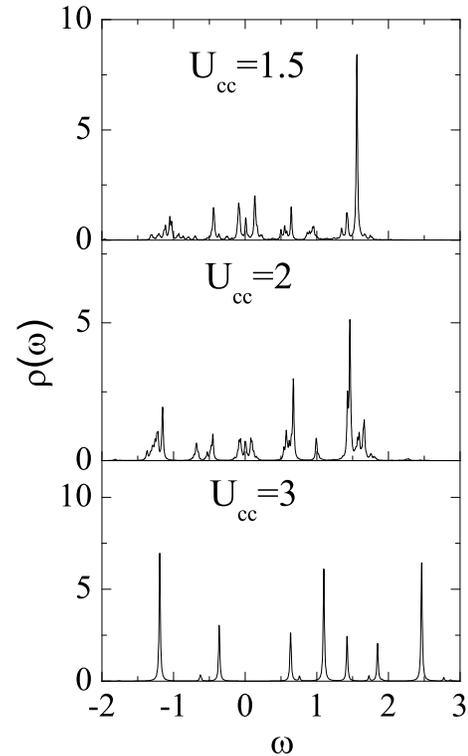}
\caption{The DOS of the light particles at third filling for
different values of $U_{cc}$ and fixed $U_{cf}=2$. ($T=0.01$,
$D=1$).} \label{fig12}
\end{figure}

\begin{figure}[t]
\includegraphics[width=0.47\textwidth]{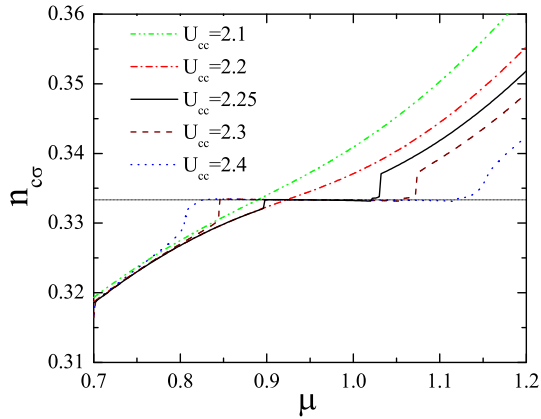}
\caption{(Color online) The  hopping particle filling
$n_{c\sigma}$ as a function of the chemical potential $\mu$ for
various values of $U_{cc}$ close to the transition value $U_{c}$
and fixed $U_{cf}=2$. The horizontal dotted line shows
$n_{c\sigma}=1/3$. ($T=0.01$, $D=1$).} \label{fig13}
\end{figure}

In this subsection we study the MIT at third  filling
$n_{c\sigma}=n_f=1/3$. The MIT is quite different in comparison
with the half filling case. It works with both the light and heavy
particles. In numerical calculations the chemical potential is
adjusted in order to maintain $n_{c\sigma}=1/3$. The two thirds
filling case $n_{c\sigma}=n_f=2/3$ is the particle-hole symmetry
of this one third filling case. We plot the  phase diagram for the
third filling case in Fig. \ref{fig14}. The value $U_c$ at the MIT
point quickly increases as $U_{cf}$ approaches to the value $D$,
while it slowly decreases as $U_{cf}$ increases. The insulating
phase is established only for sufficient large Coulomb
interactions (both $U_{cc}$ and $U_{cf}$). In the insulating phase
each lattice site is occupied by one particle of the light or
heavy particle components, and the Coulomb interactions prevent
any double occupation. The MIT can also be analyzed through the
self energy and the DOS of the light particles, as in the case of
half filling. The behaviors of the self energy in  the metallic
and insulating phases exhibit similar features as in the half
filling case. However, the features of the DOS are different. In
Fig. \ref{fig12} we plot the DOS for different values of $U_{cc}$
and fixed $U_{cf}$. One can see from Eq. (\ref{gimp}) that the
local Green function is a superposition of two subband Green
functions with nonequal weights. The Kondo peak, which appears in
the local subband DOS, is away from the Fermi energy and is
reduced.\cite{Hewson,Horvatic} Therefore, the peak that appears
at the Fermi level in the metallic phase is merely the
superposition of two local subband DOS, and maybe it is irrelevant
to the Kondo resonance. Thus, the MIT can occur at the third
filling in contrast to the single-band Hubbard model, where no MIT
is observed at third filling. In contrast to the half filling
case, we do not observe any region of coexistence of metallic and
insulating phase. However, close to the MIT, we observe a phase
separation that happens at fillings very close to the third
filling. In Fig. \ref{fig13} we plot the hopping particle filling
$n_{c\sigma}$ as a function of the chemical potential $\mu$ for
various values of $U_{cc}$ close to the phase transition point
$U_c$. One can see in Fig. \ref{fig13}, when $U_{cc}<U_{c}$, the
graphics of $n_{c\sigma}$ does not show any plateau at
$n_{c\sigma}=1/3$, which indicates the metallic phase. However, for
$U_{cc}$ close to $U_c$ in the insulator side, the graphics of
$n_{c\sigma}$ show not only the plateau, which indicates the
insulating state, but also a gap at fillings close to
$n_{c\sigma}=1/3$. Within the gap, the filling $n_{c\sigma}$ is
uncertainty. This feature indicates a phase separation at those
fillings. The MIT at third filling is commensurate. It does not
distinguish the light particles from the heavy ones. Any double
occupation must be vanished at the transition point. The MIT is
also observed at third filling in the three-component Hubbard
model.\cite{Gorelik,Inaba} One can notice that the commensurate
MIT always occurs at the commensurate fillings independently
whether the mass imbalance or the interaction anisotropy exist or
not. Working at the commensurate fillings, experiments could
detect this MIT.

\section{Conclusion}
We have studied the MIT in the three-component FKM by using the
DMFT with ED. We find the conditions for the particle fillings at
which the MIT can occur. In particular, the MIT can occur at total
filling $1 \le n \le 2$. When the total filling $1<n<2$, the MIT is species selective, and it
can occur only at half filling of the light particles
($n_{c\sigma}=1/2$). At the total filling of one third or two thirds, the
MIT can occur only at sufficient strong Coulomb interactions. This MIT works with both light and heavy particles. We also
observe the inverse MIT, i.e. the phase transition from insulator
to metal when the Coulomb interactions increase. The inverse MIT
can occur only for sufficient strong correlations between light and heavy particles, and the particle filling condition $n_{c\sigma}+n_f=1$.  We have
also studied in details the MIT at half and third fillings. At
half filling the reentrant effect of insulating phase is observed.
As the Coulomb interactions increase, the system first stays in
the insulating phase, then it becomes metallic, and finally it
goes back to the insulating phase. The first MIT is continuous,
while the second one is discontinuous. As in the single-band
Hubbard model, we also observe a finite region at the second MIT, where
both metallic and insulating phase coexist. However, in contrast
to the single-band Hubbard model, the MIT occurs together with asymmetric Kondo splitting in the spectra of the light particles. At the third filling, the MIT can occur only for
sufficient strong Coulomb interactions. We find the phase separation close to MIT point from
the insulator side. We have also constructed the phase diagram at
half and third fillings. However, in this paper we still restrict ourselves to
consider only the homogeneous phases. At low temperature one may
expect the stability of charge- (or/and spin-) density-wave
states. We leave these phases for further studies.

\section*{Acknowledgement}

This research is funded by Vietnam National Foundation
for Science and Technology Development (NAFOSTED) under grant No 103.02-2011.29.


\begin{thebibliography}{9}

\bibitem{Mott}
N. F. Mott, {\it Metal insulator transitions}, Taylor and Francis,
London (1990).

\bibitem{Bloch} I. Bloch, J. Dalibard, and W. Zwerger, Rev. Mod. Phys.
\textbf{80}, 885 (2008).

\bibitem{Jordens} R. J\"{o}rdens, N. Strohmaier, K. G\"{u}nter, H. Moritz, and
T. Esslinger, Nature (London) \textbf{455}, 204 (2008).

\bibitem{Schneider} U. Schneider, L. Hackerm\"{u}ller, S. Will,
Th. Best, I. Bloch, T. A. Costi,
R. W. Helmes, D. Rasch, and A. Rosch, Science
\textbf{322}, 1520 (2008).

\bibitem{Spiegelhalder}
F. M. Spiegelhalder, A. Trenkwalder,D. Naik, G. Hendl, F. Schreck, and R. Grimm,
Phys. Rev. Lett. \textbf{103}, 223203 (2009).

\bibitem{Taie}
S. Taie, Y. Takasu, S. Sugawa, R. Yamazaki, T. Tsujimoto,
R. Murakami, and Y. Takahashi,
Phys. Rev. Lett. \textbf{105}, 190401 (2010).

\bibitem{Gorelik}
E. V. Gorelik and N. Bl\"{u}mer, Phys. Rev. A \textbf{80}, 051602(R) (2009).

\bibitem{Inaba}
K. Inaba, S. Y. Miyatake, and S. I. Suga, Phys. Rev. A \textbf{82}, 051602(R) (2010).

\bibitem{Dao}
T.-L. Dao, M. Ferrero, P. S. Cornaglia, and M. Capone, Phys. Rev.
A \textbf{85}, 013606 (2012).

\bibitem{FreericksLemanski}
J. K. Freericks, M. M. Maska, A. Hu, T. M. Hanna, C. J. Williams,
P. S. Julienne, and R. Lemanski, Phys. Rev. A 81, 011605 (2010);
82, 039901(E) (2010).

\bibitem{Ates}
C. Ates and K. Ziegler, Phys. Rev. A \textbf{71}, 063610 (2005).

\bibitem{FKM}
L. M. Falicov and J. C. Kimball, Phys. Rev. Lett. \textbf{22}, 997 (1969).

\bibitem{Kennedy}
T. Kennedy, Rev. Math. Phys. \textbf{6}, 901 (1994).

\bibitem{Portengen}
T. Portengen, Th. \"{O}streich, and L. J. Sham, Phys. Rev. B
\textbf{54}, 17452 (1996).

\bibitem{Tran1}
Tran Minh-Tien, Phys. Rev. B \textbf{67}, 144404 (2003).

\bibitem{Tran2}
V.-N. Phan and M.-T. Tran, Phys. Rev. B \textbf{72}, 214418
(2005).

\bibitem{Fledderjohann}
U. Brandt, A. Fledderjohann, and G. H\"{u}lsenbeck, Z. Phys. B \textbf{81}, 409 (1990).

\bibitem{FreericksZlatic}
J. K. Freericks and V. Zlatic, Rev. Mod. Phys. \textbf{75}, 1333 (2003).

\bibitem{Jedrzejewski}
J. Jedrzejewski and V. Derzhko, Physica A \textbf{317} 227 (2003).

\bibitem{Hubbard}
J. Hubbard, Proc. Roy. Soc. (London) A \textbf{281}, 401 (1964).

\bibitem{Georges}
A. Georges and G. Kotliar, Phys. Rev. B \textbf{45}, 6479 (1992).

\bibitem{Rozenberg}
M. J. Rozenberg, G. Kotliar, and X. Y. Zhang, Phys. Rev. B
\textbf{49}, 10181 (1994).

\bibitem{Metzner}
W. Metzner and D. Vollhardt, Phys. Rev. Lett. \textbf{62}, 324
(1989).

\bibitem{GKKR}
A. Georges, G. Kotliar, W. Krauth, and M. J. Rozenberg,
Rev. Mod. Phys. \textbf{68}, 13 (1996).

\bibitem{Brandt}
U. Brandt and C. Mielsch, Z. Phys. B \textbf{75}, 365 (1989).


\bibitem{Si}
Q. Si, G. Kotliar, and A. Georges, Phys. Rev. B \textbf{46}, 1261 (1992).

\bibitem{Krauth}
M. Caffarel and W. Krauth, Phys. Rev. Lett. \textbf{72}, 1545 (1994).

\bibitem{Liebsch}
A. Liebsch and H. Ishida, J. Phys.: Condens. Matter \textbf{24}, 053201 (2012).

\bibitem{Hewson}
A. C. Hewson, {\em The Kondo problem to Heavy fermions}, Cambridge University Press (1997).

\bibitem{Horvatic}
B. Horvatic, D. Sokcevic, and V. Zlatic, Phys. Rev. B \textbf{36}, 675 (1987).

\end{thebibliography}
\end{document}